\documentclass[amsmath,twocolumn,superscriptaddress,showpacs,aps,prl]{revtex4}
\usepackage{bm}
\usepackage{graphics}
\begin{document}

\title{
Spin  current and polarization in impure 2D electron systems with
spin-orbit coupling}

\author{E.G. Mishchenko}
\affiliation{Lyman Laboratory, Department of Physics, Harvard
University, MA 02138}
\affiliation{L.D.
Landau Institute for Theoretical Physics, Moscow 117334, Russia}
\author{A.V. Shytov}
\affiliation{Lyman Laboratory, Department of Physics, Harvard
University, MA 02138} \affiliation{L.D. Landau Institute for
Theoretical Physics, Moscow 117334, Russia}
\author{B.I. Halperin}
\affiliation{Lyman Laboratory, Department of Physics, Harvard
University, MA 02138}
\begin{abstract}
We derive the transport equations for two-dimensional electron
systems with spin-orbit interaction and short-range
spin-independent disorder. In the limit of slow spatial variations
of the electron distribution we obtain coupled diffusion equations
for the electron density and spin. Using these equations we
calculate electric-field induced spin accumulation and spin
current in a finite-size sample for arbitrary ratio between
spin-orbit energy splitting $\Delta$ and elastic scattering rate
$\tau^{-1}$. We demonstrate that the spin-Hall conductivity
vanishes in an infinite system independent of this ratio.
\end{abstract}

\maketitle

{\it Introduction}. The subject of the novel and quickly
developing field of spintronics is the transport of electronic
spins in low-dimensional and nanoscale systems. A possibility of
coherent spin manipulation represents an ultimate goal of this
field. Typically, spin transport is strongly affected by a
coupling of spin and orbital degrees of freedom.  The influence of
the spin-orbit interaction is two-fold. The momentum relaxation
due to diffusive scattering of carriers, e.g.\ by disorder,
inevitably leads to a spin relaxation and destroys spin coherence.
On the other hand, the controlled orbital motion of carriers can
result in a coherent motion of their spins. Thus, spin-orbit
coupling is envisaged as a possible tool for spin control in
electronic devices. In particular, it is possible to generate spin
polarization and spin currents by applying electric field, the
phenomenon known as the spin-Hall effect.

Although the study of spin-Hall effect recently evolved into a
subject of intense research
\cite{MNZ,SCN,CSS,SL,HBW,Sh,SHT,R,IBM,XX,BNM,D}, the issue remains
highly controversial. Sinova {\it et al.}~\cite{SCN} have
predicted that in a clean, infinite, homogeneous 2DES the
spin-current $\hat{ j}^i_{k} = \frac{1}{4} \{ \hat \sigma_i, \hat
v_k \}$ develops a non-zero expectation value under an external
electric field ${\bf E}$. (Here $\frac{1}{2}\hat {\bm \sigma}$ and
$\hat {\bf v}$ are the operators of the electron spin and
velocity, respectively.) The spin-Hall conductivity, defined as
the ratio $\sigma_{sH}=-j^{z}_{y}/E_x$, was predicted to have a
universal value $\sigma_{sH}=\frac{e}{8\pi}$, independent of the
magnitude of the spin-orbit energy splitting $\Delta$. The effect
of impurity scattering on a spin current has been discussed in
Refs.~\cite{SL,IBM,BNM,XX}. Schliemann and Loss \cite{SL} and
Burkov {\it et al.}~\cite{BNM} found that the spin-Hall
conductivity disappears in the dirty limit $\Delta \ll \tau^{-1}$,
reaching the universal value only for sufficiently clean regime,
$\Delta \gg \tau^{-1}$. The clean regime has been analyzed by
Inoue {\it et al.}~\cite{IBM}, who argued that the spin-current
completely disappears due to vertex corrections. Recently,
Dimitrova \cite{D} obtained the universal value independent of the
relation between the spin-orbit splitting $\Delta$ and the
impurity scattering rate.

Because the spin-current is not measurable directly, its physical
meaning is obscure. In the presence of spin-orbit interaction,
electron spin is not a conserved quantity, and a spin current is
not directly related to the transport of spins. In particular,
Rashba \cite{R} demonstrated that spin current can be non-zero
even in equilibrium, as the  symmetry of an isotropic spin-orbit
Hamiltonian allows non-zero in-plane currents $j^x_y=-j^y_x \ne
0$. 
A more meaningful quantity is spin polarization (spin
accumulation) rather than a spin current. Equilibrium currents do
not lead to spin-accumulation. It remains unclear whether the
predicted {\it nonequilibrium} spin-Hall currents $j^z_y$
accumulate near sample boundaries. Bulk polarization has been
studied in both the three-dimensional \cite{LNE} and
two-dimensional \cite{E} electron systems in the electric field.

In this Letter, we develop a consistent microscopic approach to
spin transport in impure 2DES. We derive a quantum kinetic
equation which describes the evolution of a density matrix of a
non-interacting 2DES. For length scales exceeding the mean free
path, this equation reduces to a modified diffusion equation. We
then compute spin polarization and spin current in a general
situation when the finite-size system is driven out of equilibrium
by an external electric field as well as by the density gradient.
We find that the spin current actually {\it vanishes} in an
infinite system for arbitrary $\Delta \tau$.

However, in a mesoscopic conductor connected to two massive
metallic contacts, non-equilibrium spin currents $j^z_y$ flow in
the vicinity of the contacts (as shown in Fig.~1). A non-zero
spin-Hall effect can also be achieved in an infinite system by
applying a finite frequency electric field. We evaluate the ac
spin-Hall conductivity, which is maximal for a frequency of order
of the spin relaxation rate. This result is instructive for making
a connection with previous works and clarifying the 'universality'
issue of the spin-Hall conductivity.

{\it Kinetic equation}. Non-interacting electrons in an asymmetric
quantum well can be described by a single particle Hamiltonian
\begin{equation}
\label{ham} H = \frac{1}{2m} \left( {\bf p}-\frac{e{\bf
A}(t)}{c}\right)^2 + \alpha ~\hat{\bm \eta}\cdot \left({\bf
p}-\frac{e{\bf A}(t)}{c}\right) +U_i,
\end{equation}
where ${\bf p} = - i \hbar {\bm \nabla} $ is electron momentum,
$m$ is the effective mass, ${\bf A}(t)$ is a vector potential of
the uniform electric field ${\bf E}=-{\bf \dot A}/c$, and
$\hat{\bm \eta}$ is proportional to the electron spin operator.
(We neglect terms qubic in ${\bf p}$.) The disorder potential
$U_i$ is assumed to be random, short-range, and spin-independent.
For the isotropic (``Rashba'') spin-orbit interaction \cite{BR},
$\hat{\bm \eta} = {\bf z} \times \hat{\bm \sigma}$, where
$\hat{\bm \sigma}$ are the Pauli matrices. To describe a
non-equilibrium state of the system, we use the Keldysh approach
\cite{RS}. We introduce the retarded and advanced Green's
functions $G^R$ and $G^A$, and Keldysh function $G^K$ satisfying
Dyson's equation
\begin{equation}
\label{dyson} \left(\hat G^{-1}_0-\underline{\hat \Sigma} \right)
\underline{\hat G}=1, ~~
\underline{\hat G} = \left(\begin{array}{ll} \hat G^R & \hat G^K \\
0 & \hat G^A
\end{array} \right)
\ .
\end{equation}
Here the lower bar denotes  a matrix in Keldysh space,
$\hat{G}_0^{-1}=i\partial_t-\hat H$, and $\nu=\frac{m}{2\pi}$ is a
density of states per spin direction. Neglecting weak-localization
effects, one can relate the self-energy $\underline{\hat{\Sigma}}$
to the Green's function $\underline{\hat{G}}$ by a standard
disorder averaging technique \cite{AGD},
\begin{equation}
\label{scba} \underline{\hat \Sigma} = \frac{\delta_{\bf
xx'}}{2\pi\nu\tau} ~\underline{\hat G} ({\bf x},{\bf x})\ .
\end{equation}
We consider only the limit where $\tau^{-1}$ and $ \Delta$ are
small compared to the Fermi energy $p^2_F/2m$. In the absence of
electron-electron interactions, functions $\hat{G}^R$ and
$\hat{G}^A$ are independent of the non-equilibrium state  of the
system. In the Fourier representation, they are given by
\begin{equation}
\label{g-r-a} \hat{G}^{R,A}_{{\bf p}\varepsilon} =
\frac{1}{\epsilon - \xi_p - \Delta_p \,\hat{\eta}_{\bf p}
           \pm \frac{i}{2\tau}}
\ .
\end{equation}
Here $\xi_p=(p^2-p_F^2)/2m$ is the kinetic energy counted from the
equilibrium chemical potential, $\Delta_p=\alpha p$ is the energy
of the spin-orbit splitting, and $\hat \eta_{\bf p}=\hat{\bm
\eta}\cdot {\bf p}/p$ is the projection of the spin operator
$\hat{\bm \eta}$ onto the direction of the electron momentum. The
Keldysh function $\hat{G}^K$ satisfies the equation
\begin{equation}
\label{kinetic} [\hat{G}^R]^{-1} \hat{G}^K - \hat{G}^K
[\hat{G}^A]^{-1} =\hat\Sigma^K \hat G^A -\hat G^R \hat \Sigma^K.
\end{equation}

\noindent It is now customary to apply the Wigner transformation
to Eq.~(\ref{kinetic}), i.e.\ the Fourier transform with respect
to the relative time and space arguments,
\begin{eqnarray}
\hat G^K(t_+{\bf x}_+;t_- {\bf x}_-) = \frac{i}{\pi}\int
\frac{d\varepsilon d^2p}{(2\pi)^2}~\hat g_{\bf p \varepsilon}(
t,{\bf x})\nonumber\\ \times\, e^{i[{\bf p}+e{\bf A}(t)/c]\delta
{\bf x}-i\varepsilon \delta t },
\end{eqnarray}
here $t_\pm=t\pm \delta t$ and ${\bf x}_\pm={\bf x}\pm \delta {\bf
x}/2$. In the semiclassical approximation, the Wigner transform of
the right-hand side of Eq.~(\ref{kinetic})   can be replaced by a
product of the Wigner transforms of $\Sigma$ and $G$:
\begin{eqnarray}
\label{boltz} \frac{\partial \hat{g}_{\bf p\varepsilon}}{\partial
t} + \frac{1}{2}\left\{\frac{{\bf p}}{m} + \alpha {\bm \eta},
         {\widetilde \nabla}\hat{g}_{\bf p\varepsilon}\right\}
+ i \alpha [\hat{\bm \eta} \cdot {\bf p},\hat g_{\bf p \varepsilon}] = \\
= -\frac{\hat{g}_{\bf p\varepsilon}}{\tau} + \frac{i}{\tau} \left(
\hat{G}^R_{\bf p \varepsilon} \hat{\rho}_\epsilon -
\hat{\rho}_\epsilon \hat{G}^A_{\bf p \varepsilon} \right) \ ,
\nonumber
\end{eqnarray}
where ${\widetilde \nabla}= \nabla +e{\bf
E}\,{\partial}_{\varepsilon}$, and
\begin{equation}
\label{rho-definition} \hat \rho_{\varepsilon}=\frac{1}{2\pi\nu}
\int \frac{d^2 p}{(2\pi)^2} \, \hat g_{\bf p \varepsilon}
\end{equation}
is the density matrix of electrons with the energy $\varepsilon$.
The total number of particles and their total spin can be
expressed via $\hat \rho_{\varepsilon}$ as follows,
\begin{equation}
N=\text{Tr}\, \nu \!\int\! d\varepsilon \, \hat
\rho_{\varepsilon}, ~~{\bf S}=\frac{1}{2}\, \text{Tr}\, \nu\!
\int\! d\varepsilon\, \hat{\bm \sigma} \hat \rho_{\varepsilon}
\end{equation}
In the limit $\tau \to \infty$ the equation~(\ref{boltz}) reduces
to the ballistic equation of Ref.~\onlinecite{MH}. Note, however,
that the function $\hat{g}_{\bf p\epsilon}$ is not a distribution
function in the conventional sense, since it depends on both
energy and momentum.

A stationary solution to the quantum kinetic equation
(\ref{boltz}) is of the form $\hat{g}_{\bf p \varepsilon} =
A_\varepsilon (\hat{G}^R_{\bf p \varepsilon}- \hat{G}^A_{\bf p
\varepsilon})$, where $A_\varepsilon$ is an arbitrary scalar
function of the electron energy $\varepsilon$. This solution
represents the state in which the charge density is uniform, and
spin density is zero. In a non-equilibrium state with the
characteristic length scales of the spin and charge densities
exceeding the electron mean free path $l=v_F\tau$, the
distribution $\hat g_{\bf p \varepsilon}$ relaxes slowly to
equilibrium. To describe this relaxation, we derive the equation
for the density matrix $\hat{\rho}_\varepsilon ({\bf r}, t)$. It
is useful to move small gradient terms to the right hand of the
kinetic equation (\ref{boltz}), so that its left hand side
describes fast relaxation to the local equilibrium distribution:
\begin{equation}
\label{boltz-rewritten} (\partial_t+\tau^{-1}) \hat g_{\bf p
\varepsilon}+i\Delta_p [\hat{\eta}_{\bf p},\hat g_{\bf p
\varepsilon}]=\hat {\cal K}_{\bf p \varepsilon}\equiv \hat {\cal
K}^{(0)}_{\bf p \varepsilon}+\hat {\cal K}^{(1)}_{\bf p
\varepsilon}\ ,
\end{equation}
where
\begin{eqnarray}
\hat {\cal K}^{(0)}_{\bf p \varepsilon}[\hat \rho_\varepsilon]&=&
{i}{\tau}^{-1}[ \hat G^R_{\bf p \varepsilon}\hat
\rho_{\varepsilon} - \hat \rho_{\varepsilon}G^A_{\bf p
\varepsilon}]
\nonumber\\
\label{k} \hat {\cal K}^{(1)}_{\bf p \varepsilon}[\hat g_{\bf p
\varepsilon}]&=&-\frac{1}{2} \left\{  \frac{\bf p}{m} + \alpha
\hat {\bm \eta}, \widetilde \nabla \hat g_{\bf p \varepsilon}
\right\} \ .
\end{eqnarray}
Small anisotropic deviations from local equilibrium
are due to the gradient term
$\hat {\cal K}^{(1)}_{\bf p \varepsilon}$ in the kinetic equation
which can be treated perturbatively.
The solution to
Eq.~(\ref{boltz-rewritten}) can formally be written (in the Fourier
representation with respect to time) as,
\begin{eqnarray}
\label{anzats} & \hat g_{\bf p \varepsilon}&=
i\frac{(2\Delta_p^2-\Omega^2)\hat {\cal K}_{\bf p
\varepsilon}+2\Delta^2_p \hat{\eta}_{\bf p} \hat {\cal K}_{\bf p
\varepsilon}\hat{\eta}_{\bf p}-\Omega\Delta_p [\hat\eta_{\bf
p},\hat{\cal K}_{\bf
p\varepsilon}]}{\Omega(4\Delta_p^2-\Omega^2)}\nonumber\\ &&
\equiv{\cal L}[\hat {\cal K}_{\bf p \varepsilon}],
\end{eqnarray}
where $\Omega=\omega+i/\tau$. In a zeroth order, one can neglect
the gradient term $\hat {\cal K}^{(1)}_{\bf p \varepsilon}$
altogether, so that Eq.~(\ref{anzats}) gives the distribution
$\hat g^{(0)}_{\bf p \varepsilon}$ in terms of the density matrix
$\hat \rho_\varepsilon$. In the first order, we substitute the
obtained expression for $\hat g^{(0)}_{\bf p \varepsilon}$ in the
gradient term $\hat {\cal K}^{(1)}_{\bf p \varepsilon}$ to obtain
an improved expression for the distribution function, $\hat
g^{(1)}_{\bf p \varepsilon}$. This procedure is then to be
repeated to the necessary order,
\begin{eqnarray}
\hat g^{(0)}_{\bf p \varepsilon} &=& {\cal L} \left[\hat {\cal
K}^{(0)}_{\bf p \varepsilon}[\hat \rho_{\varepsilon}]\right],\nonumber \\
\hat g^{(i)}_{\bf p \varepsilon} &=& \hat g^{(i-1)}_{\bf p
\varepsilon}+{\cal L} \left[\hat {\cal K}^{(1)}_{\bf p
\varepsilon}[\hat g^{(i-1)}_{\bf p \varepsilon}]\right],~~~ i\ge
1.
\end{eqnarray}
Integrating the second-order approximation over the momentum ${\bf
p}$, one arrives at the diffusion equation for the density matrix
$\hat \rho_\varepsilon$. In a quasistationary regime ($\omega \tau
\ll 1$) the equation takes the following form:
\begin{equation}
\label{diff_eq} \frac{\partial \rho_\varepsilon}{\partial t} +
D\widetilde \nabla^2 \hat\rho_\varepsilon +i C [{\bm
\eta},\widetilde \nabla\hat\rho_\varepsilon]+B \{{\bm \eta}
,\widetilde \nabla\hat\rho_\varepsilon\}
=\frac{\hat\rho_\varepsilon}{\tau_s}- \frac{{\bm \eta}\cdot
\hat\rho_\varepsilon {\bm \eta}}{2\tau_s}. 
\end{equation}
The first two terms in this equation describe spin and charge
diffusion with $D = v_F^2 \tau / 2$ being the conventional
diffusion constant, and $v_F=p_F/m$ the Fermi velocity. The third
term describes a spin precession due to the drift velocity, and
the fourth term describes the coupling between charge and spin.
The right hand side of Eq.~(\ref{diff_eq}) describes spin
relaxation due to the Dyakonov-Perel mechanism \cite{DP}. The
coefficients of the Dyakonov-Perel spin-relaxation,  spin-density
coupling and spin-precession are, respectively,
\begin{equation}
\label{constants}
\frac{1}{\tau_s}=\frac{2\Delta\zeta}{1+4\zeta^2}, ~~
 B= \frac{\alpha\zeta^2}{1+4\zeta^2}, ~~ C=
\frac{v_F\zeta}{(1+4\zeta^2)^2},
\end{equation}
where $\Delta=\Delta_{p_F}$, and the dimensionless parameter
$\zeta=\Delta\tau$ describes relative strength of spin-orbit
coupling and disorder scattering.  In deriving
Eq.~(\ref{constants}), we assumed that the spin-orbit splitting is
small compared to Fermi energy ($\Delta \ll E_F$), while the
parameter $\zeta = \Delta \tau$ is arbitrary. (Physically, $\zeta$
represents the angle of spin precession between two consecutive
collisions.) In the case of weak spin-orbit coupling or a very
clean sample ($\zeta \ll 1$), the Dyakonov-Perel relaxation time
is large compared to the elastic mean free time $\tau_s \sim
\tau/\zeta^2 \gg \tau$ and the characteristic spin relaxation
length $\sqrt{D\tau_s}$ is large compared to the mean free path.
The spin dynamics is thus slow both in space and time and
Eq.~(\ref{diff_eq}) has a meaning of a real diffusion equation for
the coupled density and spin degrees of freedom. Note that in the
limit $\zeta \ll 1$ the spin-density coupling ($B$-term) in the
equation (\ref{diff_eq}) differs from the corresponding term in
Ref.~\cite{BNM} as a result of incorrect summation of a diffusion
ladder in Ref.~\cite{BNM}. We show below that the strength of this
coupling is crucial for the magnitude of the spin-Hall effect.

In the opposite limit, $\zeta \gg 1$, spin-relaxation is fast,
$\tau_s \sim \tau$, and occurs on a length scale of the mean free
path $l$, i.e.\ locally as compared to the system size $L\gg l$.
Spin relaxation dynamics (e.g.\ propagation of a spin-polarized
injected beam) is therefore beyond the reach of the diffusion
equation and must be studied with the kinetic equation
(\ref{boltz}). However, Eq.~(\ref{diff_eq}) can still be used to
study a steady state in which  spin polarization  changes slowly
on a scale of $l$ (which will be the case for spin-Hall
conductivity, see below). One then has to retain the terms
describing density diffusion, spin relaxation and spin-density
coupling. In the vector basis,
\begin{equation}
\label{rho-components} \hat \rho_\varepsilon=\frac{n_\varepsilon}{2}
+\hat{\bm \sigma} \cdot {\bf s}_\varepsilon,
\end{equation}
the equations (\ref{diff_eq}) are reduced to,
\begin{equation}
\widetilde \nabla^2 n_\varepsilon=0,~~~ {\bf s}_\varepsilon =
-B\tau_s ~{\bf z}\times \widetilde \nabla n_\varepsilon.
\end{equation}
Total density and spin polarization are expressed in this basis
as: $N= \nu\int d\varepsilon  n_\varepsilon$, and ${\bf S}=
\nu\int d\varepsilon  {\bf s}_\varepsilon$.

{\it Spin-accumulation}. We now apply the spin diffusion equation
(\ref{diff_eq}) to analyze spin accumulation in a finite-size
sample of the length $L$ contacted by two ideal unpolarized
metallic leads.  The sample is infinite in the transverse
direction so that $\hat\rho_\varepsilon(x)$ depends on the
longitudinal coordinate $x$ only. 
Note that the electric field in the sample enters Eq. (14) only
via $\widetilde\nabla = \nabla + e{\bf E}
\partial_{\varepsilon} $ and therefore can be eliminated by
shifting the energy as $\varepsilon \to \varepsilon+eEx$. Thus,
the electric field may be treated via the boundary conditions,
\begin{equation}
\hat\rho_{\varepsilon}(0)=F_{\varepsilon-eV},
~~~\hat\rho_{\varepsilon} (L)=F_\varepsilon,
\end{equation}
where $V=EL$ is the voltage bias between the two leads, and
$F_\varepsilon$ is the equilibrium Fermi-Dirac distribution.
Substituting the expansion (\ref{rho-components}) into
Eq.~(\ref{diff_eq}) we observe that
$s^x_\varepsilon=s^z_\varepsilon=0$. The other two equations
yield,
\begin{eqnarray}
\label{density} n_\varepsilon(x)&=&2(1-x/L)
F_{\varepsilon-eV}+2xF_\varepsilon/L, \nonumber\\
 \frac{d^2 s^y_\varepsilon}{dx^2}&-&\frac{s^y_\varepsilon}{L^2_s}
=\frac{B}{D} \frac{dn_\varepsilon}{dx}, ~~~~~ L_s^2=D\tau_s.
\end{eqnarray}
Note that the $B$-term in the equation for $n_\varepsilon$ leads
to small corrections,  $\sim \alpha^2/v_F^2$, which must be
neglected in the considered approximation. The solution to the
second of Eqs.~(\ref{density}) yields,
\begin{eqnarray}
\label{spin_density} S^y(x) = \frac{eE_\text{eff}\zeta}{2\pi v_F}
\left(1-\frac{\cosh{[\gamma-x/L_s]}}{\cosh{\gamma}}\right),
\end{eqnarray}
where $\gamma=L/2L_s$ is the dimensionless spin-flip rate, and
$E_\text{eff}=V/L$. For an infinite system, $\gamma\to \infty$,
the spin accumulation (\ref{spin_density}) agrees with the
previous calculation by Edelstein \cite{E}.

{\it Spin-current}. The spin current, as defined in the
introduction, is found from the Keldysh Green's function,
\begin{equation}
\label{sc}
 j^i_k=\frac{1}{8m} ~ \text{Tr}~ \hat \sigma^i \left(
\nabla'_k-\nabla_k
 \right)_{\bf x'\to x}\hat G^K +\frac{\alpha}{2}\epsilon_{ikz}N.
\end{equation}
The function $\hat G^K$ can be expressed via the density $\hat
\rho_\varepsilon$ with the help of the equation, $\hat G^K=\hat
G^R \hat\Sigma^K \hat G^A$, which follows from Dyson's equation
(\ref{kinetic}). After simple transformations,
\begin{eqnarray}
\label{sc11} j^i_k&=& \frac{i}{8\pi m\tau} ~\text{Tr}~ \hat
\sigma^i(\nabla'_k-\nabla_k)_{\bf x'\to x} \int d\varepsilon d{\bf
y} ~ \hat G^R_\varepsilon ({\bf x-y})\nonumber\\ && \times
\hat\rho_\varepsilon ({\bf y}) ~ \hat G^A_\varepsilon ({\bf y-x'})
+\frac{\alpha}{2} \epsilon_{ikz} N.
\end{eqnarray}
Keeping now in the integrand only the zero and first-order terms
in the expansion of $\hat \rho_\varepsilon$ over ${\bf y}-{\bf
x}$, we arrive at the final expression for the non-equilibrium
spin current in terms of the density and spin distribution
functions,
\begin{equation}
\label{spin_current} j^i_k=-e
 \frac{\delta^i_z
\zeta^2 [{\bf z}\times {\bf E_{\rm eff}}]_k}{2\pi (1+4\zeta^2)}+
\frac{v_F \zeta(\delta^i_z S^k-\delta^i_k S^z)}{1+4\zeta^2} \ .
\end{equation}
Here ${\bf E}_{\text{eff}}={\bf E} - {\bm \nabla N}/2 e\nu$ is the
gradient of the electrochemical potential including both the
electric field and the gradient of electron density.

Substituting Eq.~(\ref{spin_density}) into
Eq.~(\ref{spin_current}) we observe that the two contributions to
$j^i_k$ cancel each other in the bulk of a sample. Therefore, the
dc spin current vanishes independent of the relative strength of
disorder and spin-orbit interaction. Thus, we generalize the
result by Inoue {\it et al.} \cite{IBM} to arbitrary values of
$\zeta$. (The finite value of the spin-Hall conductivity obtained
in Ref.\ \cite{D} was due to mishandling of the electric field
vertex in the calculation of the spin-Hall conductivity.)

\begin{figure}[t]
\label{fig1}
\resizebox{.43\textwidth}{!}{\includegraphics{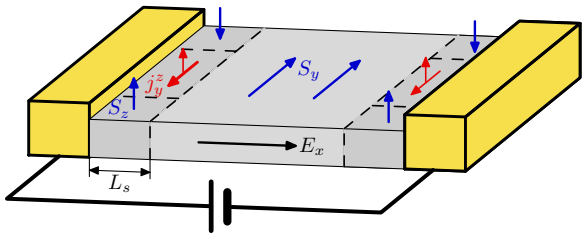}}
\caption{(color online) In a spin-Hall bar setup,  the electric
current is driven through 2DES contacted by the metallic leads
connected to a voltage source. The electric field $E_x$ creates an
in-plane spin polarization $S_y$ in the bulk. Spin currents
$j_y^z$ are running in the vicinity of the contacts while
vanishing in the bulk. Out-of-plane polarization $S_z$ is
accumulated at the sample corners. }
\end{figure}
However, near the contacts where the spin polarization deviates
from its bulk value, the spin current is non-zero. Using the
expression (\ref{spin_density}) in Eq.~(\ref{spin_current}), we
find that the spin current near the contacts decays as ($\zeta \ll
1$),
\begin{equation}
j^z_y(x)=-\frac{eE}{2\pi}~\zeta^2 e^{-{x}/{L_s}}\ .
\end{equation}
For a sample of finite width, this spin current should lead to a
spin accumulation within a distance $L_s$ of the corners of the
sample, with a component of ${\bf S}$ along $\pm {\bf z}$, as
illustrated in Fig.~1.

Note, that for a non-uniform system in thermal equilibrium, where
${\bf E}_\text{eff}= 0$, the spin density given by
Eq.~(\ref{density}) is zero, as well as the spin current. Small
equilibrium spin currents \cite{R}, proportional to
$(\alpha/v_F)^3$, are beyond the approximation used when deriving
Eq.~(\ref{spin_current}). Our derivation of the diffusion equation
(\ref{diff_eq}) and the spin-current (\ref{spin_current}) relies
on the approximation (\ref{scba}) that neglects contributions from
diagrams with crossed impurity lines (ladder approximation). This
is usually justified provided that $E_F \tau \gg 1$. In an
infinite system the result (\ref{spin_current}) is equivalent to a
calculation within the Kubo formalism with the first term
representing a single-loop contribution and the second term
originating from the ladder impurity diagrams.

To reconcile our result for spin current with the predictions of
Refs.~\cite{SCN}, it is helpful to consider the ac spin-Hall
effect~\cite{IBM}. When the applied electric field is
time-dependent, spin polarization is retarded with respect to the
field, due to the finite spin relaxation time. As a result, the
spin polarization contribution in Eq.~(\ref{spin_current}) does
not exactly cancel the electric field contribution, and spin-Hall
conductivity becomes non-zero. Solving Eq.~(\ref{boltz-rewritten})
for the homogeneous infinite system and generalizing
Eq.~(\ref{sc11}) for a time-dependent state, we find,
\begin{equation}
\sigma_{sH}(\omega) = \frac{e\Delta^2}{2\pi} ~ \frac{\omega
\tau}{\omega\tau[4\Delta^2
                    - (\omega + \frac{i}{\tau})^2]
            + 2 i \Delta^2
            }
\ .
\end{equation}
For low frequencies, $\omega \tau_s < 1$, the spin-Hall
conductivity remains small, $\sigma_{sH} \sim -i \omega \tau $.
When the frequency exceeds the spin relaxation rate ($\omega
\tau_s \geq 1$), $\sigma_{sH}$ reaches its maximum value $e \tau
/( 4\pi \tau_s)$. For clean samples, this is the universal value
$e/8\pi$ predicted in Refs.~\cite{SCN}, while for dirty samples
($\zeta \ll 1$) the maximum value of the spin-Hall conductivity
remains strongly suppressed, $\sigma_{sH}\approx e\zeta^2/(2\pi)$.

To conclude, we derived quantum kinetic equation for 2D electrons
in the presence of spin-orbit coupling and short-range potential
scattering. We proved that the dc spin Hall effect disappears in a
bulk sample, and we computed the spin accumulation in a
finite-size system for a wide range of parameters. This work was
supported by NSF grants PHY-01-17795 and DMR02-33773.

\end{document}